# Taking a Language Detour: How International Migrants Speaking a Minority Language Seek COVID-Related Information in Their Host Countries

GE GAO, University of Maryland, USA
JIAN ZHENG, University of Maryland, USA
EUN KYOUNG CHOE, University of Maryland, USA
NAOMI YAMASHITA, Kyoto University, Japan

Information seeking is crucial for people's self-care and wellbeing in times of public crises. Extensive research has investigated empirical understandings as well as technical solutions to facilitate information seeking by domestic citizens of affected regions. However, limited knowledge is established to support international migrants who need to survive a crisis in their host countries. The current paper presents an interview study with two cohorts of Chinese migrants living in Japan (N=14) and the United States (N=14). Participants reflected on their information seeking experiences during the COVID pandemic. The reflection was supplemented by two weeks of self-tracking where participants maintained records of their COVID-related information seeking practice. Our data indicated that participants often took *language detours*, or visits to Mandarin resources for information about the COVID outbreak in their host countries. They also made strategic use of the Mandarin information to perform selective reading, cross-checking, and contextualized interpretation of COVID-related information in Japanese or English. While such practices enhanced participants' perceived effectiveness of COVID-related information gathering and sensemaking, they disadvantaged people through sometimes incognizant ways. Further, participants lacked the awareness or preference to review migrant-oriented information that was issued by the host country's public authorities despite its availability. Building upon these findings, we discussed solutions to improve international migrants' COVID-related information seeking in their non-native language and cultural environment. We advocated inclusive crisis infrastructures that would engage people with diverse levels of local language fluency, information literacy, and experience in leveraging public services.

CCS Concepts: • **Human-centered computing** → **Human computer interaction (HCI);** *Empirical studies in HCI*

**Additional Key Words and Phrases:** Information seeking, crisis, migrants, minority language, inclusion

**ACM Reference format:**
Ge Gao, Jian Zheng, Eun Kyoung Choe, and Naomi Yamashita. 2022. Taking a Language Detour: How International Migrants Speaking a Minority Language Seek COVID-Related Information in Their Host Countries. *Proc. ACM Hum.-Comput. Interact,* 6, CSCW2, Article 542 (November 2022), 32 pages. https://doi.org/10.1145/3555600

542

---

This work is in part supported by the National Science Foundation, under grant 2147292 and grant 1753452.
Author's addresses: Ge Gao, Jian Zheng, and Eun Kyoung Choe, University of Maryland, College of Information Studies, 4139 Campus Drive, College Park, Maryland 20742, USA, [gegao, jzheng23, choe]@umd.edu; Naomi Yamashita, Kyoto University, Department of Social Informatics, Yoshidahonmachi, Sakyo Ward, Kyoto, 606-8317, Japan, naomiy@acm.org.

# 1 INTRODUCTION

Information seeking is crucial for people to survive a crisis and the aftermath [9, 26, 30, 74, 96]. In the field of HCI and CSCW, scholars have dedicated substantial efforts to establishing timely and reliable connections between crisis information and the affected public. Much of their work explores how people gather information from news media and/or personal networks to estimate the evolving risk of an ongoing crisis [41, 42, 58, 82, 98], stay connected with other stakeholders responding to the crisis [27, 45, 46, 98, 107], combat the threat of misleading content [2, 59, 63, 66], and restore regular life routines as well as mental health [57, 71, 88, 108]. Despite its values, this research has mostly been conducted with domestic citizens of the affected regions. It leaves international migrants, who have to navigate the crisis outside their native language and cultural environment, in a near blind spot.

Our review of the broader literature found a small set of studies examining international migrants' life experience and mental status in times of crises. They outline a concerning situation. For example, Sato and coauthors surveyed 230 Brazilians living in Niigata, Japan. Their data showed that 67% of the respondents lacked the essential crisis information when the Mid-Niigata Earthquake occurred. People either did not know there were shelters to go or could not discern the exact locations of the shelters nearby [87]. Burke and coauthors conducted focus groups with 21 Latino migrants living in northern California, the United States. Participants reported that they felt vulnerable during hurricanes because there were no local radio programs or alert systems designed for them to follow [15]. A more recent study by Yen and colleagues interviewed a total number of 60 Chinese, Italians, and Iranians living in the United Kingdom during the COVID pandemic. Interviewees described themselves as short of both informational and social support compared to the locals [106]. Together, these studies point to the fact that crisis information infrastructures in a host country often do not serve the international migrants' needs adequately. To ensure the life quality of everyone affected by a crisis, research should investigate how international migrants seek crisis information and why information resources designed for the domestic citizens fail them.

The current paper presents the first step toward understanding and designing for international migrants' crisis information seeking in their host countries. Specifically, we conducted a qualitative study with two cohorts of Chinese migrants, one living in Japan and the other in the United States during the COVID pandemic. Participants provided retrospections on their COVID-related information seeking behaviors, motivations, and concerns while living abroad. They also maintained self-tracking records of their ongoing information seeking practices over two weeks. From the above processes, we obtained a rich dataset that consisted of various comparison groups (e.g., Japan versus United States) and data streams (i.e., interview responses and self-tracking records), enabling triangulation. This study design prepared us for a close examination of participants' crisis information seeking in the context of the COVID outbreak.

In summary, we found that the Chinese migrants in our sample frequently visited Mandarin information resources, such as newsfeeds of Chinese social media platforms and conversations with fellow Chinese migrants, to gather information about the COVID outbreak in Japan or the United States. We used *language detour* to refer to this practice, as it challenges the assumption that people living in a given country will primarily count on information generated in that country's majority language to stay informed about the crisis around them.

We uncovered multiple situational factors that jointly contributed to the emergence of language detour. We also investigated the relationship between language detour and participants' consumption of local crisis information in the majority language. Notably, participants leveraged Mandarin information to scaffold their consumption of COVID-related news in Japanese or English. The potential risk of such practices was they tending to miss high-stakes information when it received insufficient or delayed coverage in the Mandarin information world. Although the risk often happened, participants were incognizant of and/or had little control over it.

Our findings expand existing knowledge of crisis information seeking by demonstrating international migrants' practices, strategies, and struggles in COVID-related information seeking. In the specific case of our sample, the Chinese migrants' information needs were not fully satisfied by either the Mandarin resources or the local resources in Japanese or English. Insights gained from the current study shed light on the future design of inclusive crisis infrastructures in times of the COVID pandemic and beyond. In particular, we outline design suggestions that facilitate crisis information seeking by international migrants with diverse levels of local language fluency, information literacy, and experience in leveraging local public services.

## 2 RELATED WORK

In this section, we review three lines of prior work that set the scene for the current research. We first summarize multidisciplinary literature on information seeking in times of public crises. It provides empirical responses to two high-level inquiries: why public crises can often trigger people's information needs and what actions people may take to satisfy such needs. We then focus on recent studies and practices in the context of the COVID pandemic. Here, we identify primary challenges people may encounter when navigating information on the pandemic. It also outlines existing solutions to facilitate COVID-related information seeking. Finally, we elaborate on the notion of international migrants as a marginalized group. We describe findings from a limited, yet indispensable, set of studies indicating the vulnerability of international migrants who experience public crises in their host countries.

### 2.1 Information Seeking in Times of Public Crises

Information serves as the basis for people to interpret the world they live in and make adequate decisions. Prior work in crisis informatics and risk communication has proposed various theoretical models explaining the major drivers of information seeking (e.g., [1, 38, 39, 53, 54]). Despite differences, all the models draw a direct connection between information seeking and perceived uncertainty. For example, the Risk Information Seeking and Processing Model posits that people are motivated to collect additional information when their current knowledge is insufficient vis-à-vis what they need to know about a given situation [37]. The Theory of Motivated Information Management conceptualizes information seeking as a person's primary means to cope with the anxiety arising from excessive uncertainty [1]. Further studies differentiate between internal and external uncertainty: The former refers to the inability to make judgments and decisions from a layman's perspective [5], whereas the latter emphasizes a lack of scientific knowledge about the objective circumstance [78].

When people perceive uncertainty in times of public crises, they collect information by making deliberate choices among various sources. Previous HCI and CSCW research documents many such cases. Specifically, Palen and coauthors examined peer-to-peer communication during multiple natural disasters in the 2000s. They found that people often referred to the information exchanged within their intimate networks (e.g., family members, neighbors in the same hazardous area) for evacuation planning [75, 90]. Gui and colleagues conducted a series of studies during the Zika virus outbreak between 2015 and 2017. Their data and analyses demonstrated how English speakers leveraged Reddit as an English-dominant medium for collective information seeking and sensemaking of the novel virus [41, 42, 59]. The bulk of literature has investigated crisis information seeking via popular social media sites in the affected countries, such as Twitter in the United States and Facebook in Germany (e.g., [58, 82]). It portrayed a fast-paced information flow connecting multiple stakeholders, including the affected public [6, 45], frontline responders and local governments [46, 107], and other individuals and agencies who were willing to help [27, 90]. Emerging from this body of literature is the belief that the access to more information enhances people's confidence to resolve uncertainty and survive a crisis. But is this always true?

## 2.2 Challenges in COVID-Related Information Seeking and Their Solutions

Recent research on the COVID pandemic suggests that consuming crisis information will sometimes increase a person's feeling of uncertainty rather than reducing it. While studies conducted during other crises have suggested the similar (e.g., [42, 44, 82]), the COVID pandemic arguably creates a more chaotic situation where both the demand for information and the challenges experienced by information seekers are amplified. Much literature in this space highlights the problem of mis-and-disinformation and its consequence. Lack of expert consensus [83, 99], eagerness to make sense of the unknown [55, 110], and political intention to create and blame some "responsibly opponents" [2] all contribute to the prolific false narratives about COVID. As a result, information seekers have sometimes felt confused about what risk measures or prevention guidelines they should follow [77, 110]. Other studies have examined the issue of information overload. They found that people felt overwhelmed not only because of the huge volume of COVID-related information accumulated [61, 113], but also due to the ambiguity and novelty of that information [81], leading to the perception of having too much to process [91].

The above challenges are likely to shift people's long-term information seeking practices toward somewhat problematic directions. For example, Kim and colleagues analyzed quantitative survey responses from participants in South Korea, Singapore, and the United States. They found that a person's previous exposure to COVID-related false information resulted in their intentional avoidance of further information regardless of its quality [55]. This finding has been verified by survey studies conducted in several other countries [43, 91]. Furthermore, Pine and coauthors conducted in-depth interviews with participants in the United States. They discovered that most interviewees tended to favor personal contacts over the mass media when seeking COVID-related information after the initial episode of the outbreak [77]. Such a practice enabled people to pinpoint information that is most trustworthy in accordance with their personal criteria, but it also risked trapping the information seeker inside an echo chamber [112].

Local governments and research communities have been endeavoring to offer the public accurate and pertinent information through ICT-based services. National COVID data dashboards, for instance, constitute one common format of presenting official and authoritative information (e.g., [25, 102]). A growing amount of visualization research and tools further advance this effort by enabling data comprehension specific to the information seeker's personal interest [29, 64, 84, 86, 111]. Moreover, many countries have implemented mobile-based contact tracing apps and symptom checkers for ego-centered risk management. Examples of the apps include Germany's Corona-Warn-App [24], Poland's ProteGO Safe [79], and Italy's Immuni [48]. While these services appear to enhance people's crisis information seeking, most of them, if not all, focus on information seekers who are domestic citizens of the affected country. Other populations, such as international migrants living in that country, and their characteristics are often downplayed in the design of crisis information infrastructures (see [20] for an exception).

### 2.3 International Migrants as a Marginalized Group

International migrants have received very limited coverage in crisis information seeking research. Nevertheless, existing evidence suggests that the available information resources may not benefit domestic citizens and international migrants living in an affected country equally [40, 72, 73].

Specifically, there have been a handful of studies revealing how language barriers prevent international migrants from accessing information in their host countries. Bhandari and coauthors interviewed 14 Nepalese migrants working or studying in Japan during the COVID outbreak in mid-2020. Most interviewees reported that they lacked sufficient Japanese proficiency to leverage local websites and interpersonal communication for healthcare information [8]. Yen and colleagues' study, conducted within a similar time frame, involved multiple groups of international migrants living in the United Kingdom. Participants sometimes found it difficult to comprehend COVID-related information in English [106]. Maldonado and coauthors reviewed all 47 websites of Council of Europe government ministries. Only 6% of those websites provided information on COVID testing or healthcare entitlements in common foreign languages, which puts international migrants who are in a language minority at a clear disadvantage [67].

Several other studies highlight the fact that international migrants are often unfamiliar with the host country's information infrastructure. For example, Chang and colleagues' research suggested that it could take international students a long time to adopt the social media and other digital systems used in their host countries [17, 18]. Chu and colleagues examined Chinese and Hungarian students' strategies of information searching via English websites. Participants in their studies encountered many difficulties in identifying the appropriate websites for each searching task. As a result, they reported feeling clueless about the quality of the searching outcomes [22, 85]. Ito surveyed 196 Vietnamese people living in Japan during the COVID outbreak. Most of the survey responders lacked precise information about the social welfare packages they already had; they did not know where to seek that information either [50]. Such a finding echoes those reported by earlier studies of multiple immigrant groups living in the United States [14, 69, 92].

Another thread of research examines the social challenges experienced by international migrants during the COVID pandemic. While this research does not directly speak to international migrants' information seeking practices, it indicates that the local network they may consult for information and support is usually small. For instance, Misirlis and coauthors surveyed 248 international students in the Netherlands. Participants perceived a lack of connection with local communities in the host country, which contributed to their feeling of depression during the pandemic [68]. Lai and colleagues surveyed 124 international students who came to the United States and the United Kingdom prior to the pandemic. They compared participants who stayed in their host countries after local COVID outbreaks with those who returned to their home countries. The stayers experienced significantly higher anxiety, and they reached out to local contacts less for informational and emotional support [62]. Ang and Nancy's research reported that the spread of COVID in Singapore elicited local residents' negative attitude toward international people living there [3]. Adding to these findings, some qualitative studies have documented cases where international migrants distanced themselves from the locals to avoid conflicts [101, 106]. For instance, multiple Chinese and Italian participants in Yen et al.'s research reported that their local colleagues or family members in the United Kingdom thought overacted to COVID. To cope with the conflicts, participants sometimes chose to stay away from COVID-related conversations and media consumption in the workplace or at home [106].

## 3 RESEARCH QUESTIONS

Putting up different groups of literature together, we see questions that are important but remain unanswerable by previous research. In particular, there is little research outlining the landscape of international migrants' information seeking in times of crises. Existing evidence suggests that individuals should be able to gain local crisis information from many resources, ranging from word-of-mouth communication to official notifications issued by the government. However, most of this evidence was collected from the affected region's domestic citizens exclusively. Several studies incorporating international migrants indicated there were information resources they could not fully leverage, but there lacks empirical understanding of people's current or conceivable practices addressing challenges in crisis information seeking. Examples of those challenges include low-quality information and information overload.

We argue that establishing knowledge in the above space is crucial for HCI researchers and practitioners: It will enable us to identify what our target population strives to achieve and where existing approaches fall behind; it will also inspire technology design assisting all those affected, which, by definition, includes people who experience the crisis in a non-native language and cultural environment. The ongoing COVID pandemic provided a situation that enabled us to investigate such knowledge. To that end, we asked the following research questions (RQ):

*RQ1*: Where do international migrants seek COVID-related information in their host countries? Why do they prioritize those information resources?

*RQ2*: What challenges or risks do international migrants perceive with COVID-related information seeking in their host countries? What workarounds have they developed in reaction to these challenges or risks, if any?

The rest of this paper presents a qualitative study that answers the above RQs.

## 4 METHOD

Our research process began with in-depth interviews with 28 Chinese migrants living in Japan and the United States. These interviews took place in August and September 2020, during COVID outbreaks in these two countries. From the data collection phase and onwards, we iterated through generating codes and concepts from collected data and elaborated on them by referring to additional data as well as related literature. We stopped the data analysis after clear connections emerged among all the core concepts. The leading investigators for data collection and analyses are native Mandarin speakers who have previous or ongoing experiences living in Japan and the United States. Such a team composition enables us to be sensitive to the contextual meanings of participants' self-reports. Below, we present further details of our research method.

### 4.1 Sampling Strategy and Recruitment Process

The research team brainstormed sampling criteria that would best allow us to understand international migrants' information seeking practices during the COVID pandemic. This brainstorming resulted in three parameters of our research design:
- *Shared language background between participants and the research team*: Participants and researchers should speak one or more shared native languages. Empirical and anecdotal evidence both suggest that international migrants often feel most comfortable detailing thoughts and feelings in their native languages. Thus, our research design should provide participants the option of doing so;
- *Involvement of participants in multiple host countries:* Participants should be recruited from more than one host country, enabling open comparisons between different groups of participants. In the context of our study, we hoped to leverage these comparisons to detect the association, if any, between participants' information seeking practices and the cultural environment in each particular host country;
- *Availability of migrant-oriented information resources at the host countries:* Ideally, the host countries involved in our research should have COVID-related information resources already set up for international migrants. This condition would allow us to examine the extent to which our participants were aware of and utilized migrant-oriented resources in addition to others, such as their contacts or social media platforms.

These parameters led to our decision to recruit Chinese migrants living in Japan and the United States. We started the recruitment process by posting digital flyers on popular discussion forums among Chinese migrants in both countries. Qualified participants were people who a) moved to the host country from mainland China, b) remained in the host country during the COVID pandemic, and c) felt comfortable communicating with the researchers in either Mandarin or the majority language of the host country (i.e., Japanese or English). Our final sample consisted of 28 participants: half lived in Japan during the period of our study and half in the United States. Their demographic information is described in Table 1.

Table. 1. Demographic information of all participants.

| ID | Host Country | Duration | Current Status of Occupation | Level of Proficiency in the Host Country's Majority Language |
|---|---|---|---|---|
| C-JP 1 | Japan | 2 years | Graduate student, with part-time jobs | Professional working proficiency |
| C-JP 2 | Japan | 2 years | Graduate student | Limited working proficiency |
| C-JP 3 | Japan | 2 years | Graduate student | Elementary proficiency |
| C-JP 4 | Japan | 6 years | Graduate student | Professional working proficiency |
| C-JP 5 | Japan | 6 years | Graduate student | Professional working proficiency |
| C-JP 6 | Japan | 2 years | Graduate student | Elementary proficiency |
| C-JP 7 | Japan | 2 years | Graduate student | Limited working proficiency |
| C-JP 8 | Japan | 2 years | Graduate student | Limited working proficiency |
| C-JP 9 | Japan | 1 year | Graduate student, with part-time jobs | Limited working proficiency |
| C-JP 10 | Japan | 1 year | Graduate student, with part-time jobs | Elementary proficiency |
| C-JP 11 | Japan | 2 years | Graduate student | Professional working proficiency |
| C-JP 12 | Japan | 3 years | Graduate student, with part-time jobs | Professional working proficiency |
| C-JP 13 | Japan | 2 years | Graduate student | Limited working proficiency |
| C-JP 14 | Japan | 2 years | Graduate student | Elementary proficiency |
| C-US 1 | United States | 3 years | Graduate student | Professional working proficiency |
| C-US 2 | United States | 1 year | Graduate student | Elementary proficiency |
| C-US 3 | United States | 11 years | Full-time employee | Professional working proficiency |
| C-US 4 | United States | 1 year | Graduate student | Professional working proficiency |

| C-US 5 | United States | 1 year | Graduate student | Limited working proficiency |
| --- | --- | --- | --- | --- |
| C-US 6 | United States | 4 years | Undergraduate student | Professional working proficiency |
| C-US 7 | United States | 1 year | Graduate student | Professional working proficiency |
| C-US 8 | United States | 1 year | Graduate student | Professional working proficiency |
| C-US 9 | United States | 1 year | Graduate student | Limited working proficiency |
| C-US 10 | United States | 6 years | Undergraduate student | Professional working proficiency |
| C-US 11 | United States | 3 years | Ful-time employee | Professional working proficiency |
| C-US 12 | United States | 6 years | Graduate student, with part-time jobs | Professional working proficiency |
| C-US 13 | United States | 5 years | Graduate student | Limited working proficiency |
| C-US 14 | United States | 2 years | Graduate student | Professional working proficiency |

*  **Note**: Participants described their Japanese or English proficiency following definitions given in the ILR scale [49].

## 4.2 Data Collection

We collected the main body of our research data through qualitative methods. Specifically, each participant first attended a semi-structured interview with us. This interview included questions about a) the interviewee's overall experience of living in the host country, b) any changes that the local COVID outbreak had brought to their lives, c) the online and/or offline resources that they often utilized for COVID-related information seeking, and d) their reflections on whether and why knowing that information would matter to their lives in the host country. All the interviews were conducted over the online conferencing platforms preferred by the participant; each lasted for about one hour. We audio recorded these interviews with participants' consent.

Upon the completion of the first interview, participants performed self-tracking of their COVID-related information seeking practices for two weeks so as to collect detailed incidents that triangulated the interview data. We provided a mobile-based tracking app, implemented using OmniTrack for Research [56], to assist participants in the tracking process. The app sent out four prompts daily to participants' mobile phones at 9am, 1pm, 5pm, and 9pm. All the prompts contained an identical set of questions, asking the participants to report a) the most important COVID-related information they learned about over the past 4 hours, if any, b) the social interactions they had with other people over the past 4 hours, if any, and c) their perceived level of anxiety at the moment (4-point Likert scales adopted from Tluczek et al. [93]; 1 = not at all, 4 = very much). Participants could provide their responses using various forms of inputs, ranging from textual descriptions to screenshots of the information they had consumed. They could also choose to skip any of the prompts or specific questions included in a prompt that asked for their COVID-related information seeking or social interactions. We made the deliberate choice to design our self-tracking instructions in this way so that participants would not feel they had to consume COVID-related information or interact with others for being involved in the study.

By the end of the self-tracking period, we reached out to each participant again for a retrospective interview. Participants were asked to review a randomly selected subset of their tracking record. They reflected on their motivation, process, and consequence of seeking each piece of information in the record. Similar to the first interview, this post-tracking interview also happened over online conferencing platforms preferred by the participants. Each session lasted for 30-40 minutes, and it was audio recorded with participants' consent.

Our final dataset consisted of interview responses and self-tracking records in multiple languages. Although participants were informed that they could choose Mandarin, Japanese, or English for their interviews, all of them communicated with the researchers in Mandarin. Self-tracking records contained information in a mixture of languages, including textual descriptions generated in Mandarin and screenshots of online information in all three languages. Participants living in Japan responded to 2.5 prompts sent from the self-tracking app per person per day (SD = 0.47), and they detailed two pieces of COVID-related information per person in every three days. Participants living in the United States responded to 2.9 prompts per person per day (SD = 0.50), and they detailed one piece of COVID-with information per person in every two days.

### 4.3 Analysis

We performed inductive analysis of the interview responses and the self-tracking records. Following suggestions from Glaser and Strauss [34, 35], we iterated through collecting data, generating new codes from collected data, and checking and elaborating on the codes by collecting more data until saturation point was reach, that is, no new codes were generated. This grounded theory approach has commonalities with thematic analysis of certain types (e.g., codebook thematic analysis, reflective thematic analysis; [12, 23]) as they can all help the researcher discover patterns of meanings from often unstructured data. Nevertheless, Glaser and Strauss recommend a distinct multiphase procedure for developing qualitative theories of the phenomenon of interest [34], which aligns well with the ultimate goal of our research project.

In the open coding phase, the first and second authors of this paper read through all the available data independently and developed codes to exhaust a randomly selected batch of this data. They then compared and discussed the codes before repeating the same process with subsequent data batches as the data collection continued. We generated 298 codes and 1507 quotations by the end of the open coding. These codes and quotations captured participants' practices of COVID-related information seeking, their interpretations of and preferences among different information resources, their perceived challenges or risks of crisis information seeking in the host country, and current strategies to work around them.

In the axial coding phase, we worked on pinpointing the relationships among codes and concepts generated during open coding. We leveraged three sources of information to ensure the credibility of our relationship interpretation: logic connections and reasonings explicitly stated by participants, triangulation among relevant data of different formats, and the theoretical connections between concepts as implied by previous research. This information enabled us to further categorize the initially scattered codes. It also assisted us in connecting data of different categories into a reasonable structure.

In the selective coding phase, we identified language detour as the central category in response to our research inquires. We then iterated again through refining the existing codes and categories and revisiting relevant literature for the purpose of outlining a coherent model connecting other main categories to the central one. The following section details our findings about participants' crisis information seeking featuring language detour.

## 5 FINDINGS

Our data suggested that participants heavily relied on Mandarin information to learn about the COVID outbreak in their host countries. Frequently mentioned sources of the information include social media platforms that use Mandarin as their default operational language and conversations with fellow Chinese migrants. We use language detour to refer to this practice, as it counters the assumption that people primarily count on information generated in the host country's majority language, predominant on local sources, to stay informed about updates around them. Table 2 and Table 3 summarize how language detour was evident in participants' interview responses and self-tracking records, respectively.

Much of the Mandarin information described the spread of COVID at the country level. For example, participants regularly read news reports produced by Chinese media to learn about the status of COVID outbreak in Japan and the United States. These reports usually appeared in Sina Weibo's Trending Topics board, with the content tailored to each reader partially based on their location. Participants also reported visiting certain business accounts via WeChat and Sina Weibo to check detailed descriptions of COVID-related breaking news (e.g., newly announced regulations; hereinafter referred to as *detailed information*) as well as the daily COVID cases number update (hereinafter referred to as *number updates*) at their host country. Two frequently mentioned business accounts were Dealmoon (北美省钱快报) and 1Point3Acres (一亩三分地). The former is an online shopping platform listing recent deals in North America; the latter is a discussion forum for information exchange about academic programs and job opportunities globally. Both services were initially established by and for Chinese migrants living abroad. Since spring 2020, they incorporated a module of summarizing COVID-related detailed information and number updates in various countries.

Table. 2. Top 10 information resources identified in participants' interview responses when describing how they sought information about the COVID outbreak in their host countries.

| Information Resource and the Percentage of Participants Who Relied on it to Seek Information About the COVID Outbreak in Japan | | | Information Resource and the Percentage of Participants Who Relied on it to Seek Information About the COVID Outbreak in the United States | | |
|---|---|---|---|---|---|
| Platform and Language | Information Stream | Perc. | Platform and Language | Information Stream | Perc. |
| WeChat (Mandarin) | Feed of daily updates (i.e., WeChat Moments) generated by all WeChat accounts a person has followed | 100% | WeChat (Mandarin) | Feed of daily updates (i.e., WeChat Moments) generated by all the WeChat accounts a person has followed | 100% |
| WeChat or in-person, individual (Mandarin) | One-on-one conversations with a fellow Chinese migrant | 78.6% | WeChat or in-person, individual (Mandarin) | One-on-one conversations with a fellow Chinese migrant | 92.9% |
| Sina Weibo (Mandarin) | Feed of news (i.e., Trending Topics) recommended by the platform, and feed of daily updates from all the Weibo accounts a person has followed | 64.3% | Sina Weibo (Mandarin) | Feed of news (i.e., Trending Topics) recommended by the platform, and feed of daily updates from all the Weibo accounts a person has followed | 78.6% |
| WeChat or in-person, group (Mandarin) | Group conversations among multiple fellow Chinese migrants | 57.1% | Local institutional emails (English) | Official notice issued by a person's local university or company | 57.1% |
| Yahoo! Japan (Japanese) | Feed of news recommended by the platform | 50.0% | Zhihu (Mandarin) | Feed of questions and answers (Q&As) recommended by the platform | 42.9% |
| LINE News (Japanese) | Updates on the daily number of new COVID cases in Japan | 42.9% | WeChat or in-person, group (Mandarin) | Group conversations among multiple fellow Chinese migrants | 35.7% |
| Local hospital websites (Japanese) | Official notice issued by the hospital | 42.9% | Twitter News (English) | Updates on the daily number of new COVID cases in the United States | 35.7% |
| Local institutional emails (Japanese) | Official notice issued by a person's local university or company | 42.9% | Wall Street Journal mobile app (English) | Feed of news recommended by the platform | 21.4% |
| NHK News (Japanese) | Feed of news recommended by the platform | 35.7% | Local hospital websites (English) | Official notice issued by the hospital | 21.4% |
| Zhihu (Mandarin) | Feed of Q&As recommended by the platform | 28.6% | John Hopkins Univ. website (English) | Information on the COVID resource center webpage | 21.4% |

* Note: Among all the listed platforms, WeChat, Sina Weibo, and Zhihu originate in China. WeChat and Sina Weibo are two social networking sites. Zhihu is a Q&A site. The default operational language of these three platforms is Mandarin.

Table. 3. The language and content distribution of participants' self-tracked information seeking about the local COVID outbreak in their host countries.

| Language Type[y] | Percentage of Self-Tracked Information Seeking About the COVID Outbreak in Japan | | Percentage of Self-Tracked Information Seeking About the COVID Outbreak in the United States | |
|---|---|---|---|---|
| | Detailed Information | Number Updates | Detailed Information | Number Updates |
| Mandarin | 26.2% | 9.3% | 44.4% | 9.7% |
| Japanese | 25.6% | 39.0% | - | - |
| English | - | - | 33.3% | 12.5% |
| Total | 100% | | 100% | |

Another subset of information in Mandarin helped our participants remain aware of what was happening in the nearby geospatial areas. For example, many participants identified the personal updates posted by their WeChat and Sina Weibo friends as an essential source to learn if "*someone working in the grocery store next to our apartment building just got tested positive,*" or "*the local pharmacy had sold out of protective masks.*" Several of the participants also mentioned they joined self-organized WeChat groups of Chinese migrants in the same county or groups managed by the Chinese Students and Scholars Association at their university. Members of these groups periodically broadcasted the information they had, such as the locations and capacity of nearby COVID testing sites, to the rest group members. Participants interpreted the above information as a sensitive indicator showing the severity level of the COVID outbreak in their local areas. Local, in this context, usually referred to the district or the neighborhood a person was living in.

In the rest of this section, we elaborate on three aspects of our findings. They jointly form a comprehensive view of how international migrants in our study performed COVID-related information seeking through language detours. We first identify salient characteristics of participants' life environments that prompt them to turn to Mandarin information. We then present scenarios where participants leverage Mandarin information as a strategic device to gather and make sense of COVID-related information provided in the host country's majority languages (i.e., Japanese or English). We conclude this section with participants' reflections on their unique identity as crisis information seekers standing in between two language worlds. For the clarity of presentation, we use italics within quotation marks to indicate direct quotations from participants' interview responses.

It should be noted that our data indicated no systematic association between participants' information seeking practices and the environment specific to each host country. Thus, we do not differentiate between participants living in Japan or the United States throughout the majority of our findings.

## 5.1 Mandarin Information as a Window into the Local COVID Outbreak

Participants in our study leveraged Mandarin information as their window into the local COVID outbreak. A significant part of our interview data revealed reasons leading to this practice. Surprisingly, we did not observe clear associations between participants' reasons for taking language detours and their self-reported level of proficiency in Japanese or English. Instead, there appeared to be specific characteristics of the migrants' life under COVID which prompted people with even high proficiency in the local languages to perform much of their information seeking in Mandarin. We detail these characteristics below.

*5.1.1 Fast Comprehension with only Fragmented Time.* Most participants explicitly stated that they have used fragments of time to collect information about the local COVID outbreak. Despite the daily increase in COVID cases, life went on. Participants who were international students experienced the heavy burdens of remote learning. Many of them reported that they signed up for more courses than usual because the COVID outbreak elicited their worries about "*next steps in this country if the pandemic continues.*" Gaining more knowledge was perceived by these participants as an actionable means to better prepare for an unforeseeable future. Similarly, those employed in part- and full-time jobs worked harder for the financial security of themselves and their families under an especially challenging time. COVID-related information seeking mostly happened when participants were "*taking a lunch break at work,*" "*trying to refresh my mind in between two classes,*" and "*checking my cellphone at the end of an exhausting day.*"

Reading COVID-related information in Mandarin enabled our participants to speedily comprehend what had been happening. While these participants had no problem studying or working in the host country's majority language, they preferred Mandarin information for efficiency. As explained by C-US 12:

> "*Everything I read in school and at work is in English. I don't know if this is because of the pandemic or whatnot. I just feel there is too much information to process. I need a mental break in my personal time. Mandarin information works better for this purpose because it takes little effort to process. I can finish reading a Mandarin news article in maybe thirty seconds. If the same article was written in English, I don't know, I might have to spend a few minutes on reading it.*" [C-US 12]

Similarly, participants living in Japan also tried to avoid reading long pieces of news in Japanese. Nevertheless, several of them paid close attention to news headlines written in Japanese. The Mandarin characters (kanji) embedded in those headlines helped these Chinese migrants quickly grasp the high-level meaning of the news without reading into the details:

> "*Most of the time, I go to WeChat or Sina Weibo. There are Mandarin posts that describe important stuff going on in Japan. The only places for me to read Japanese news are the TVs in on-campus dining halls. I glance over the titles of the Japanese news. It's good enough for me to gain the basic ideas because I can recognize COVID-related keywords. For example, Mandarin (感染) and Japanese (感染) use the same characters for the word 'affection.' City names in Japanese are mostly written in Mandarin characters too.*" [C-JP 9]

*5.1.2 Convenient Access via Unmarked Platforms.* All participants in our study suspected that domestic citizens in their host countries relied on local media for COVID-related information. However, most participants could not specify what those local media were, with only a couple of exceptions (i.e., NHK News in Japan; the Wall Street Journal and CNN in the United States). Participants frequently visited social media platforms that originate in China, such as WeChat and Sina Weibo, for online information seeking. This choice brought them into an information world where the vast majority of the content was created in Mandarin. Here, we adopt Trubetzkoy and Jakobson's notion of "markedness" [95] to categorize the roles that different platforms played in our participants' information seeking practices. Chinese social media sites and apps constituted unmarked platforms for our participants, given people's self-reports that they used those sites and apps as "*the most convenient and natural information sources.*" In contrast, local platforms used by domestic citizens formed a group of marked and, in many cases, obscure information sources to participants in our study.

The interview data revealed elements that contributed to participants' preference for Chinese social media platforms over local media. Most participants reported that they had never purchased TVs for their living spaces. Only three people across our entire sample owned TVs in the apartments they rented, but there were no local cable channels connected to the devices. Beneath such a life condition was these migrants' shared feeling that they were "*constantly moving between different short-term leases*" and "*unsettled.*" Many of them had the experience of watching local TV news in public dining halls or the gym spaces of their apartment buildings. However, this limited exposure was "*not enough for me to be familiar with most of the local media.*"

A few of our participants regularly visited social media platforms featured in their host countries for COVID-related information. Despite that, they encountered considerable difficulty in identifying the exact media or journalist accounts to follow, given a cultural environment that was different from their home one. Participants reported that they consulted the local information resources mainly for breakdowns of the COVID number updates in Japan and the United States. Chinese social media sites and apps, with their default operational language of Mandarin, remained the unmarked platforms for our participants to learn detailed information, such as COVID prevention strategies and breaking news in their host countries. As one example, C-US 6 explained her problems with information seeking via Twitter, which echoed other participants' experiences of using LINE in Japan:

> "*I guess many locals in the United States use Twitter to learn about what just happened. I check Twitter too, but I don't really know much about this platform. I don't know which media accounts or which person's accounts to follow there. If you showed me the Sina Weibo accounts of two Chinese news reporters, I would be able to tell, 'okay, I should go to this person for entertainment news, but the other one for political news.' I wouldn't be able to recognize such differences among reporters in the United States.*" [C-US 6]

*5.1.3 Word-of-Mouth Within a Shrunken Network.* Our interview data indicated interpersonal conversations as another primary source for participants to collected COVID-related information. However, only a tiny proportion of those conversations involved domestic citizens of the host countries. Table 4 summarizes participants' self-tracking records regarding their social interactions with others, which confirmed the pattern that emerged from the interview responses.

It is noteworthy that participants in our study reported frequent interactions with local schoolmates and colleagues before COVID. Besides attending classes or working together, people made friends with each other when "*joining the same sports clubs,*" "*hanging out at the social hours and picnics organized by the college,*" and "*grabbing food together after work.*" For many participants, local contacts used to serve as their sources to learn about the essential social and informational infrastructures in their host country.

While none of the participants intended to decrease their interactions with local people, they all identified the implementation of remote work guidelines as a turning point in their social lives. In particular, the guidelines required individuals and institutions to either cancel all the in-person activities or transfer those activities to a virtual format. An unexpected consequence, experienced universally among participants, was that they "*suddenly found it extremely difficult to have casual conversations with those who were not Chinese.*" C-US 14 and C-JP 11 elaborated on their experiences in this regard:

> "*I was in a dance club before the pandemic. I made friends with local people there, and we talked quite often. After the COVID outbreak, all events at the club were canceled, and people organized online hangouts instead. I attended once. They chatted about stuff like the American TV shows they watched, but I didn't share those memories. It was just stressful, like I was attending a TOEFL test. I never went to those social hours again. Recently I joined an online book club of Chinese international students. That one is much better.*" [C-US 14]

> "*I have been talking online with my Japanese lab mates after our campus closed. But the conversations are all about work. It would be awkward to schedule a meeting just to talk about life. We always start with work-related matters, and it is not easy to switch the topic because people use honorifics for formal conversations in Japanese. There is always this feeling of distance. I mean, we talk, but there is a tacit agreement that we will not bring personal matters into the conversation.*" [C-JP 11]

Based on the above quotations and similar ones from other participants, we infer that performing joint activities in a collocated space is essential for many international migrants to establish spontaneous conversations with their local contacts. In contrast, online interactions appeared challenging because there lacked convenient clues for people with different language and cultural backgrounds to identify shared topics and build intimacy.

With the absence of local contacts, participants exhausted their network of fellow migrants to satisfy their social and informational needs. They set up a rich variety of social activities in small groups, including *"hosting online cocktail parties,"* *"playing murder mystery games over their phones,"* and *"forming remote reading clubs."* These activities not only served as the participants' venues for having fun, but they also enabled information exchange among individuals who faced similar anxieties elicited by the pandemic. It came as no surprise that Mandarin was the dominant language at such activities.

Table. 4. The contact and topic distribution of participants' self-tracked social interactions with others living in their host countries.

| Contacts Identity | Percentage of Self-Tracked Interactions with Others Living in Japan | | Percentage of Self-Tracked Interactions with Others Living in the United States | |
|---|---|---|---|---|
| | Work Conversations | Personal Chitchats | Work Conversations | Personal Chitchats |
| Chinese Migrants | 6.7% | 65.3% | 13.1% | 74.0% |
| Japanese Locals | 14.7% | 2.1% | - | - |
| American Locals | - | - | 2.9% | 2.7% |
| Other Internationals | 8.6% | 2.6% | 4.4% | 2.1% |
| Total | 100% | | 100% | |

*5.1.4 Positive Thinking by Reading Between the Lines.* Last but not least, our participants described information seeking in Mandarin as an opportunity to develop positive thinking about the local COVID outbreak.

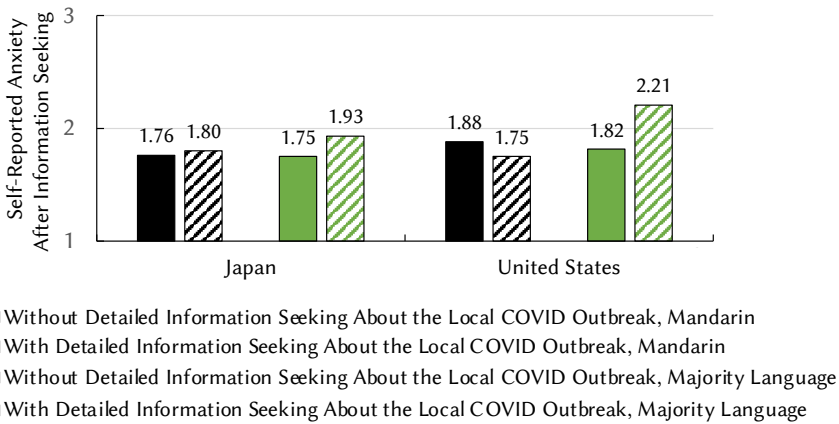

Fig. 1. Participants' self-reported anxiety after seeking information about the COVID outbreak in their host countries.

Many shared an emotional trajectory where they *"got super upset by the local COVID outbreak at the beginning,"* responded by *"searching COVID-related information every day for good news and hopes,"* then gradually moved to *"feeling helpless because the pandemic just continues."* When these participants read Mandarin information, such as COVID-related news posts via WeChat or Sina Weibo, they looked for *"sentences that encouraged the readers to stay strong"* and *"humorous or sarcastic expressions that helped tone down the threat of COVID."* However, none of the participants reported the same strategy when describing information seeking in Japanese or English.

Participants' self-tracking records of their perceived anxiety partially confirmed the findings from interview responses. As illustrated in Figure 1, participants reported a higher level of anxiety when they had read detailed descriptions of the local COVID outbreak written in Japanese or English (i.e., labeled as "with detailed information seeking"), as opposed to reading the local COVID number updates only (i.e., labeled as "without detailed information seeking"). Reading detailed descriptions written in Mandarin did not elicit additional anxiety from our participants.

### 5.2 Multilingual Scaffolds for COVID-Related Information Seeking

The heavy consumption of Mandarin information did not make Mandarin the exclusive language of a participant's information world. In particular, our data captured cases where participants made strategic use of Mandarin information to work around the challenges or potential risks they perceived with crisis information seeking in Japanese or English. The rest of this section describes two dominant patterns in which these international migrants moved back and forth between COVID-related information of the same topics but across multiple languages. While such practices enhanced participants' perceived effectiveness of COVID-related information gathering and sensemaking, they disadvantaged people through sometimes incognizant ways.

*5.2.1 Screening, Scrutinizing.* One common practice mentioned by all participants was a two-step workflow of information seeking. During the first step, participants browsed a wide range of COVID-related information written in Mandarin. They relied on this step not only to *"keep up with as many updates as possible"* but also to *"filter out information that was not worth careful reading."* As a follow-up step, they turned to Google Search or the local media quoted in the Mandarin information. The ability to comprehend multiple languages allowed these participants to *"delve into corresponding but selected reports"* written in Japanese or English.

Further, our analysis suggests that participants were especially motivated to collect multilingual descriptions of the same piece of information when this information was pertinent to their lives and/or open to alternative viewpoints:

*Pertinent to migrants' lives.* The host country's COVID-related international travel restrictions, visa extension requirements during boarder closure periods, and the local public's attitudes toward Chinese migrants were three clusters of information frequently discussed by participants during their interviews. When browsing news updates through WeChat or Sina Weibo, participants actively looked for information about these above topics, hoping to answer questions that *"mattered a ton to us [migrants], but little to the majority of the locals."* Examples of such questions included when they could plan for their next reunion with families in China, how long they would be able to maintain legal status in Japan or the United States, and where they could find a discrimination-free community to start their next housing lease. Participants made the deliberate choice of tracing back to source content written in Japanese or English when there was any. By doing so, they minimized the risks of having important information *"lost or delayed during the paraphrasing across languages."* Figure 2 presents one example documented in C-US5's self-tracking record. This participant leveraged Mandarin information to screen COVID-related news of interest to her, then moved to the source article in English for close reading.

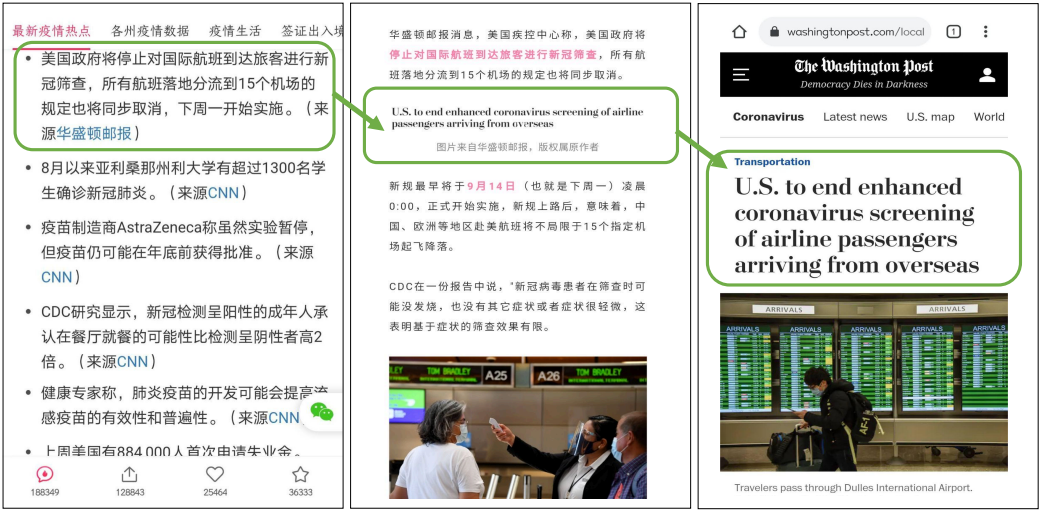

Fig. 2. An example of the "screening, scrutinizing" process of COVID-related information seeking, as reported by C-US 5. In this example, the participant followed Dealmoon's business account via WeChat. From browsing the newsfeed of this account, she noticed the headline that airports in the United States were going to end enhanced COVID screening of airline passengers from overseas (left). Paraphrases of the original news report in Mandarin were also provided (middle). The participant then moved to The Washington Post's official website, seeking details of the newly announced travel policy (right).

*Open to alternative viewpoints*. When participants described their experiences of COVID-related information seeking, they intentionally separated the factual aspect of this information from viewpoints about it. The former consisted of objective measures of COVID spread (e.g., the local number updates), public events (e.g., the launch of new projects on vaccine research and testing), and formal rules and regulations (e.g., telework policies). The latter refers to opinions produced by people who authored or reposted the information (e.g., the author's comments regarding how COVID may affect the local economy in the long term). Participants leveraged their multilingual abilities to collect alternative viewpoints of the same information if they sensed that the opinion provided in the Mandarin report was limited or biased. Such a practice was largely shaped by participants' regular exposure to the culture and public narratives of more than one country. As C-US 9 and C-JP 4 pointed out:

"*When I see eye-catching news from my Chinese friends' posts or the Sina Weibo newsfeed, I usually search it in English for full context. This is not because I doubt the facts described by the Chinese media. I don't think they would give a wrong number of the COVID cases in the United States. If they did so, people would soon figure it out. However, it is so common that different governments make contrasting interpretations of the same fact. I want to see all perspectives from both sides so that I won't be misled by either one.*" [C-US 9]

"*I can read articles written by local people [in Japan] and by Chinese authors. My sense is that everyone attempts to say their own country is in good control of the COVID spread. If the daily number is high, they would emphasize something else promising, such as the* *country is well prepared for monitoring the increase [in positive cases]. But they don't sugarcoat and probably exaggerate things when talking about the situation in another country. I've learned that it is helpful to see different versions of the story.*" [C-JP 4]

While the workflow of "screening in Mandarin" and "scrutinizing in Japanese or English" enabled our participants to balance the breadth and depth of COVID-related information seeking, it raises potential concerns. For instance, multiple participants living in Japan mentioned that the Japanese government had set up a special funding program to support international students during the COVID pandemic. However, participants were unaware of this program until they ran into WeChat or Sina Weibo news describing it. While performing a follow-up search of Japanese information sources, participants realized that "*the university had actually emailed all of us about the funding program for multiple times, but we somehow just missed it.*" A similar experience was also reported by participants in the United States when recalling how they found out about updated procedures for COVID testing at local sites.

Interestingly, those people who disregarded the crisis updates and reports in institutional emails also claimed a routine of paying close attention to assignment-relevant emails (e.g., course announcements posted by the instructor, meeting notifications sent by a project manager) from their universities or workplaces. This claim confirmed the crucial yet complex role of a screening step in international migrants' information seeking process. Specifically, participants of our study seemed to selectively read Japanese or English emails for efficiency. Being assignment-relevant was the primary filter participants applied in their navigation among a high volume of institutional emails under a regular setting. When it came to the time of the pandemic, people tended to leverage crisis information in Mandarin as an additional filter to keep track of institutional emails that were not about tasks-to-perform but as well worth careful reading. There appeared to be potential cost if participants overly relied on updates and reports written in Mandarin at the screening pass. When a piece of crisis information is reported in Japanese or English only, people are likely to miss the information despite its importance.

*5.2.2 Knowing What, Knowing How.* Another common practice revealed by our data concerned how participants drew upon multilingual information to make time-sensitive and high-stakes decisions. Representative examples of those decisions included how to schedule medical appointments after self-observing COVID symptoms and what health protocols to follow at upcoming domestic travels. Participants in both Japan and the United States reported that they paid close attention to official instructions written in their host country's majority language. However, they often encountered some knowing-doing gaps, or "*challenges of applying the overall guidelines to the migrants' practice.*" To bridge these gaps, participants turned to various Mandarin resources for additional information that was tailored to their needs.

Closer inspection of the interview data revealed issues that impeded participants from acting upon Japanese or English announcements directly. One issue was difficulty in comprehending complex words and expressions outside the context of academic writing. While participants often encountered such words, they also acknowledged checking online dictionaries and performing Google Searches as two effective means to make sense of the words.

For the majority of our participants, what upset them more was the impression that "*people who designed the instructions and websites [hosting the instructions] seemed to assume that all readers have shared certain prior knowledge or preferences.*" Participants recalled cases contradicting this assumption. The example below was reported by C-JP 10 who tried to follow the local hospital's instructions on COVID testing in Japan. Despite good comprehension of all the information written in Japanese, he found the instructions put him in a difficult situation:

> "*At some point I worried that I might have been infected with the virus. I tried to schedule a COVID test at the nearby hospital. After reading their website, I realized that people here were not allowed to schedule COVID tests by themselves. You need to see a doctor first, then schedule the test only with the doctor's permission. I've seen doctors in Japan once or twice but, still, it feels a bit uneasy. I'm not quite familiar with medical procedure here. If I said anything wrong, it might cause me trouble during this sensitive time. The requirement [of seeing the doctor first] makes sense for the locals, but not for people like me.*" [C-JP 10]

The above participant ended up getting a COVID test after collecting more actionable guidance in Mandarin. This guidance came from two primary sources: COVID testing experiences and tips shared in Chinese migrants' WeChat groups, and the Embassy's webpage listing local hospitals that specialized in serving international visitors. Other participants living in the United States also mentioned the similar practice of leveraging Mandarin information to interpret local COVID-related instructions. The interpretation work requires information providers to offer additional understandings and insights satisfying the migrants' needs, which goes far beyond simple language translation.

Notably, participants perceived a few status differences between "know-what in Japanese and English" and "know-how in Mandarin". For many of them, instructions written in the host country's majority language carried "*a sense of legitimacy.*" It would be the only information for everybody to count on "*especially when dealing with misunderstandings and filing disputes.*" The Mandarin information could contribute useful understanding but was considered as being informal. Besides legitimacy, sustainability surfaced as another focus of people's attention. Participants were concerned that they might not always be able to find sufficient Mandarin guidance for COVID-related actions, especially those at a local level.

**5.3  Standing In-Between Two Information Worlds**

In addition to speaking about specific scenarios and practices, participants reflected on their unique identity as crisis information seekers standing in between two worlds. One of these worlds contained rich information generated "*in the majority's language and mainly for the locals of the host country.*" The other was the Mandarin world that "*involves some but not all the information relevant to us living abroad.*" During the COVID pandemic, participants perceived that they were in a marginalized position compared with domestic citizens of their host countries as well as their home country. This perception was closely associated with attributes of the crisis information available in each world.

In particular, our participants benefited little from migrant-oriented information resources that had been set up by the local governments and public facilities of their host countries. Most of these people could not recall whether any of the public authorities' websites had offered multilingual content. Only four participants had visited Mandarin pages under their local governments' websites, such as the Japan National Tourism Organization's (JNTO) and the Centers for Disease Control and Prevention's (CDC). While those pages were designed to provide COVID-related information for international migrants living in Japan or the United States, participants found the content there "*highly summary,*" "*disfluent,*" and "*not quite informative.*"

The Mandarin information provided by Chinese social media platforms was, in general, perceived to be comprehensible and instructive. However, it required participants to "*handpick content*" pertaining to COVID-related life tips and policies in Japan or the United States. For example, many online articles in Mandarin advocated taking traditional Chinese medicines for COVID prevention. This information did not benefit but instead disconcerted our participants, as they "*could not purchase those medicines at local pharmacies.*"

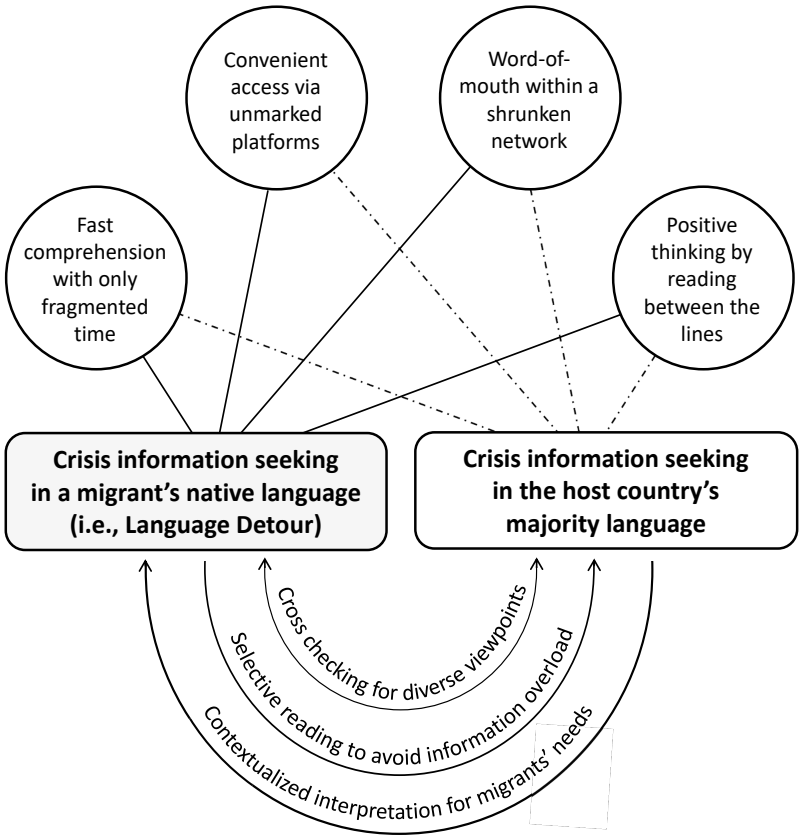

Fig. 3. A model that describes how international migrants perform crisis information seeking in their host countries whose language and cultural environments are different from the person's native one. In particular, this practice features language detour where a migrant consumes information generated in their native language to stay informed about and prepare for the crisis in a host country.

## 6 DISCUSSION

The overarching goal of this current study was to understand and enhance international migrants' crisis information seeking in their host countries. To this end, we interviewed and gathered information seeking records of Chinese migrants living in Japan and the United States during the COVID outbreak. Our data indicated that participants in both countries preformed language detour, the practice of leveraging Mandarin information to stay informed about and prepared for their local COVID situation (see Figure 3 for illustration). Further analysis found language detour a double-edged sword, and it did not relieve participants from feeling marginalized. In the following discussion, we elaborate on how a migrant's COVID-related information seeking can be shaped by three conditions of their life in a host country: fluency in the majority language, information literacy within local cultural context, and experience in leveraging local public services. We conclude each section with corresponding insights and opportunities for inclusive design of crisis information infrastructures.

### 6.1 Situated Aspects of Language Fluency

Previous research suggested that language barriers often impede international migrants from making use of crisis information resources in their host countries. Thereout, they highlighted the necessity to assist migrants' information seeking with translation services (e.g., [40, 76]). In that body of literature, an information seeker's language fluency was often conceptualized as roughly binary: either the person would have no problem understanding crisis information generated in the given language or they would not be able to comprehend it at all. Findings of the current study challenged this binary view by demonstrating the situated aspects of language fluency and its role in shaping the participants' COVID-related information seeking.

As indicated in our findings, international migrants who came to the host country as students or expatriates were able to attend school and work in the host country's majority language. They also had adequate fluency to hold conversations with the locals and read news from local media, although it may have required help in terms of accommodations from other communicants [97] or explanations offered by a dictionary [31]. During the COVID pandemic, the capability to navigate information in the majority language proved crucial for a couple of reasons. Participants in our sample, for instance, valued information in Japanese or English as an accountable reference to understand local regulations of work, transportation, and healthcare in the time of COVID. Some of them visited featured local media for COVID number updates that were more detailed than the country level. On a relevant note, prior studies also emphasized the importance of local information resources in aligning different stakeholders for effective crisis response [27, 45, 107].

Then, why would it still be necessary to seek crisis information in a migrant's native language? Our data demonstrated that this practice was partially a natural consequence shaped by the situation and partially a deliberate choice made by the information seeker. In particular, sustained COVID transmission in a host country created unusual constraints on the Chinese migrants' information seeking in the majority language. Participants often found it too overwhelming to fully consume COVID-related updates in their limited time. They desired positively framed information to cope with anxieties elicited by the COVID but were not sensitive to the affective meanings conveyed in Japanese or English. A similar finding was also reported by CSCW research examining conflict management and self-disclosure involving multilingual speakers (e.g., [32, 65]). The amount of cognitive as well as emotional effort involved in reading COVID-related information went significantly beyond that required in regular activities (e.g., attending a meeting in Japanese or English). This extra cost pushed international migrants toward taking a language detour where they could gain a better control over not only the comprehension of crisis information but also its influence on a person's mental status.

More interestingly, participants were alert to the challenges or risks of COVID-related information seeking in a single language, no matter which language it was. They developed strategies to maximize the benefits of their multilingual ability, such as using crisis information in Mandarin as a filter to direct the consumption of Japanese or English information under corresponding topics. A potential concern of the above strategy, though, is that the information of different language versions is usually not equivalent in its quantity and/or quality. While using native-language information as a filter would prevent migrants from information overload in the majority language, it could lead to undesired ripple effects (e.g., selection bias).

*Design opportunities.* Building upon the above reflections, we argue that there needs to be ways to alleviate an international migrant's burden of consuming crisis information in a non-native language. The most straightforward approach may be to translate all the information produced in the host country's majority language into a person's native language. However, our findings indicated that such translations would not serve the best interests of a migrant nor be necessary. We advocate an alternative approach that harnesses international migrants' multilingual ability for efficient crisis information seeking in their host countries.

Specifically, our approach considers using selective translation to assist and improve the two-step process of screening-and-scrutinizing. As described by participants, separating crisis information seeking into two successive steps enabled them to quickly identify the information of interest, as the screening often happened in people's native language (e.g., Mandarin). The identified information then received careful reading in its source language (e.g., Japanese, English) so that the authenticity remained. Inspired by this practice, we propose designing a multilingual interface that translates the host country's local news headlines and briefs into the information seeker's native language. People could leverage these translations to quickly grasp the main point and decide whether they would like to read the full details of a given piece. Those details, however, would be better presented in the source language to avoid loss and bias in translation; or they could be overlayed with auto-generated paraphrases, in similar formats to explanations offered by a monolingual dictionary, to assist comprehension of complex words.

In cases where a reader can barely speak the host country's majority language, all content would have to be translated but with annotations. The annotations should present data pertaining to the accountability of the translated content, such as who generated the translations (e.g., professional translators, bilingual volunteers, or machine translation; [33]) and which parts of the translation are likely to be erroneous (e.g., the error types; [103]). Despite possible variations in the exact amount of translation needed by the information seeker, we caution against using translation as a reckless substitute for the original text.

## 6.2 Information Literacy Within a Local Context

Apart from language fluency, information literacy emerged as another factor closely associated with international migrants' crisis information seeking in the current research. A significant proportion of the social media platforms (e.g., WeChat) and information producers (e.g., business accounts, interest groups, individual influencers) mentioned by our participants were embedded in Chinese as a person's native cultural context. These information resources constituted a buffer zone facilitating international migrants' transition to a new environment in the host country.

Several recent studies have identified the sources and patterns of COVID-related information seeking by domestic residents of an affected country. A comparison between their research findings and ours manifested the uniqueness of international migrants' practice with further details. In a national survey conducted in the summer of 2020, for example, United States' residents identified traditional media, such as radio and television, as their most important venues of crisis information seeking [61]. Zhang and coauthors' longitudinal research, running from September to December 2020, not only verified this finding but also found that traditional media played a longer-lasting role in local people's coping with the pandemic than other venues (e.g., Facebook, Twitter) [110]. In contrast, when participants in our study commented on their interaction with local news broadcast on radio or TV, they pointed out a clear distance between that information world and themselves as outsiders. Besides, domestic residents sometimes avoided browsing local media websites because they were highly aware of the bias inherent in specific platforms or accounts [77, 110]. While our participants also consumed limited local media, this limitation was primarily due to unfamiliarity instead of being a deliberate choice backed by understanding.

Crisis news and updates in a person's native language (e.g., Mandarin) often direct people back to an information world of greater comfort and less uncertainty. Nevertheless, our findings cautioned that the convenience of leveraging native-language resources often came with a cost. For instance, it requires additional processes for local events in Japan or the United States to be reported in Mandarin. Over the course of these processes, the timeliness and the granularity of the source information may both decrease. When browsing casual information, such as commercials and entertainment news, timeliness and granularity is seldom a critical need [9]. However, transferring this mindset to crisis information seeking can disrupt high-stakes decision making. The last point was evident in our participants' experiences when they overlooked COVID-related support in the host country due to its delayed coverage on the Chinese media.

In the long term, there is likely a negative reinforcement loop connecting language detour and local information literacy. The more heavily a migrant relies on information resources in their native language for information seeking, the less proficient they will be in navigating and evaluating information provided by featured local resources and vice versa. Although participants' self-reported data did not directly state it, there is reason to believe that overdependence on the Mandarin information would postpone our participants' acculturation process in general.

*Design opportunities.* It takes time for international migrants to develop sufficient information literacy in their host country (e.g.,[16]). Thus, we urge the necessity to support their crisis information seeking with long-term educational efforts. Specifically, schools, universities, and workplaces should consider developing training materials and programs to foster international members' information literacy within a local context during the crisis. Our current research, along with prior work [17, 72], indicated several potential focuses of this training. For instance, many international migrants would gain from learning systematic knowledge of essential local information resources. In the case of COVID, these resources may include the COVID dashboards from local communities at various levels of geographical granulation and the COVID contact tracing services introduced by local public health authorities. Also, the training should walk attendees through the heterogeneity of viewpoints and narratives among different information producers within the host country (e.g., prefectural vs. national authorities in Japan, CNN vs. Fox News in the United States). Our current findings indicated a frequent lack of this knowledge among international migrants. Although the Chinese participants intentionally collected and cross-checked alternative versions of the same crisis information, they tended to view the information resources in Japan or the United States as a whole against those produced in the Chinese media. As a result, people were likely to miss subtle yet essential clues when calibrating their quality assessment of the crisis information.

Further, we envision several formats of technology design with a shared goal of increasing international migrants' daily exposure to the host country's local media and platforms. They complement the development of educational programs by providing more immediate assistance to the migrants. One possible design is to ask online platforms operated in a migrant's native language and cultural context (e.g., Sina Weibo, Zhihu) to display some shortcuts. The shortcuts would function as location-based recommendations pointing a user to top-visited information resources in their country or area of living. This design would enable international migrants to access timely and legitimate local updates as a byproduct of their regular information seeking activities. One practical concern, though, is the uncertainty in establishing a coordination protocol among information platforms across countries, especially when the service is for profits.

An alternative design for immediate assistance is to leverage the technique of crosslingual information retrieval. Information seekers who are international migrants could search for crisis updates in their native language and among media platforms familiar to them. The system would then return information mapped to the person's initial search query but retrieved from a pool of mixed languages and sources, similar to the function embedded in some earlier versions of the Google Search (Figure 4; [36]). Following a more sophisticated approach, if feasible, the system may automatically identify a migrant's topics of interest from their history of visiting unmarked platforms. Identified topics would then serve as a guide to prioritize local crisis updates for the migrant's reading.

In either case, we argue that it is essential to present crosslingual information with a carefully designed structure, such as displaying crisis updates of a migrant's core interest at the top or showing alternative narratives of the same event in adjacent positions. Having such a structure would prevent international migrants from being overwhelmed by the information retrieved from more than one language and cultural context. It would also enhance people's current practice of comparing multiple pieces of information that revolved around the same crisis update but containing various viewpoints and/or levels of details.

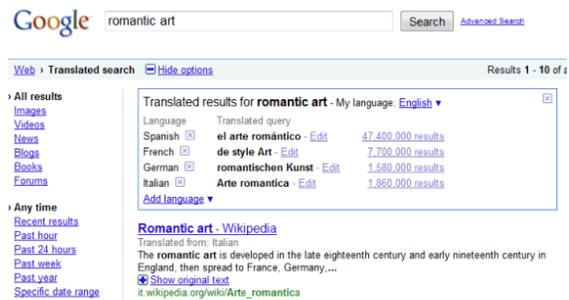

Fig. 4. An example of Google Search's crosslingual interface in 2009. Users generate search queries in one language. The search engine will then return a collection of crosslingual results based on the user's preference.

### 6.3 Migrant-Oriented Content in Consideration of a Different Living Experience

Assisting international migrants with crisis information seeking goes beyond (re)considering in which language the information should be provided and on which platforms the information should be hosted. The current research calls attention to people's living experience in a society that differs from their home. Many of our participants needed additional guidance to understand how public facilities worked in their host country. All of them had to constantly calibrate their work and life plans in according to COVID-related changes to visa policies as well as public attitudes toward the international people. However, crisis information that speaks to these people's concerns appeared sporadic or not quite explanatory.

Our findings provide hints regarding where the migrant-oriented content would initially be from and how this content could be disseminated. Specifically, participants followed several social media accounts, such as Dealmoon and 1Point3Acres, that are established by and for Chinese who live abroad. They also sought informational and emotional support from fellow Chinese migrants in their personal networks. Language detours via these Mandarin resources created benefits that local information resources were not yet able to afford. The current research specified two examples of such benefits: easy navigations among COVID-related information tailored to the Chinese migrants' needs, and contextualized interpretations that connect COVID-related guidelines for the local public with practices of the Chinese migrants.

Despite values, participants in our study were concerned over the current content generation model as it was primarily based on voluntary contributions. There was little systematic effort to gather individual migrant's wisdom; nor had people formed established ways to transform their wisdom into organized knowledge for dissemination. Moreover, while multiple participants reported their experiences of visiting migrant-oriented discussion forums or online groups for information, few people identified themselves as knowledge contributors donating their own tips on crisis coping. This dynamic echoes the repeated findings of prior research under other voluntary-based scenarios of information and support sharing (see Kraut and Resnick's [60] collection of work as one example). To date, promoting voluntary contributions for the public good still remains an open question for CSCW scholars and practitioners. Our participants expressed concerns regarding the sustainability, as well as quality of, crisis information tailored to international migrants' needs. However, they could not specify solutions except exhausting their network of fellow migrants and thinking twice before adopting takeaways shared by others.

Further, it is worth noting that the endeavor to generate migrant-oriented crisis information does not and should not all come from migrants, given their status as an already disadvantaged population. A small but growing body of recent literature has investigated the local government's role in shaping international migrants' quality of life under the COVID pandemic. Most of this research targeted issues with policy making, such as local laws defining migrants' rights in a crisis context (e.g., [13, 80, 100]) and government-driven actions affecting migrants' health and vaccination equity (e.g., [7, 28, 70]). Findings of the current research also indicated local governments' significant potential to serve international migrants with financial and informational support intended for them. Unfortunately, this top-down support, as reported by our participants, often turns out to be disconnected from the migrants' information seeking and life practices. A possible reason may be that there lacks channels and mechanisms to include international migrants' first-person accounts in public authorities' logistics of support design.

*Design opportunities*. We propose that the value of the migrant-oriented content will be maximized with properly implemented audit processes. This proposal is grounded in the fact that the migrant-oriented content described by our participants was primarily user-generated or summarized by third-party agents whose information selection procedure was not transparent. Prior research has demonstrated the feasibility of combining human and algorithmic efforts to detect misinformative content spread over social media [10, 47, 51, 89], commercial campaigns hidden among other posts on knowledge sharing sites [19], the lack of diverse information sources returned by search engines [94], and disruptive online behaviors issued of individual users [4, 21]. While most of this research was conducted in English, it developed techniques that can possibly be applied to safeguard the quality of migrant-oriented content in other languages.

Moreover, our findings point to the need for technical interventions that supervise a migrant's disclosure of their know-how to others and for others' benefit. Much of this know-how was highly experiential by nature: People generated practical knowledge based on specific trial-and-errors when navigating a crisis as international migrants; their experiences and tips were most helpful for fellow migrants who were in a comparable situation as to where they have been. This finding echoes previous observations on online peer-support communities involving marginalized groups (e.g., [104, 105]). A design space worth exploring here concerns how the information should be formulated before it reaches others. For example, the platform may assign a writing template that nudges information contributors to examine the experiences and tips they are about to share and ensure they have presented certain details. Doing so will, in turn, enhance a person's assessment of how and to which extent someone else's know-how could be transferred to their own situation.

Last but not least, we advocate ways to connect migrant-oriented content contributed by migrants themselves to that issued by public authorities so that the strengths of both parties could complement each other. In the case of COVID-related webpage design by JNTO or CDC, for instance, the government's current effort in providing multilingual content was not appreciated by our participants because it failed to satisfy migrants' information needs beyond language comprehension. Previous research has indicated that international migrants may obtain social and informational support from online communities consisting of their geographically dispersed in-group members (e.g., [109]). Our current work suggests a complementary perspective to the above one. Specifically, we suspect that people would benefit from certified knowledge archives or Q&A pages contextualizing official crisis-coping guidelines for the target audience's situations and practices. To that end, some online platforms, such as WeChat's public groups and the discussion forum of 1Point3Acres among Chinese, could already provide authentic and up-to-date references to the questions of concern of information seekers. Local public authorities should consider harnessing the migrants' thoughts and perspectives for crisis infrastructure designed to serve them. This process would generate migrant-oriented information that is more accessible and of higher accountability than that authored by individuals.

## 7 LIMITATIONS AND FUTURE DIRECTIONS

While the current research contributes an in-depth and one of the first accounts of international migrants' COVID-related information seeking experience and practice, the findings are qualified by limitations. Specifically, we chose to study Chinese migrants living in Japan and the United States based on a deliberate reflection on our positionality as researchers. This sampling strategy led us to a setting where participants could usually be connected with a social and informational network of in-group fellows. Future research should pay attention to migrants with a smaller-size fellow network in the host country or having more diverse backgrounds (e.g., occupations, educational levels). Knowledge gained from that research will deepen the understanding of how the crisis information seeking model outlined in the current work could be transferred to other scenarios of concern.

Also, we look forward to follow-up research examining international migrants' crisis coping using quantitative and unobtrusive methods. Those methods hold the promise of uncovering factors shaping people's information seeking practices or the consequences of language detour that may not be captured by qualitative interviews for various reasons. They also risk less from introducing any priming effects to the data collection process. In the current research, we have designed our interview questions as well as the self-tracking instructions carefully so that participants would not shift their natural behaviors or responses to questions for the researcher's interest. Our data confirmed that participants did follow their own choices to skip some of the prompts and consume information in a variety of formats. That said, follow-up research can help cross-validate our work by adopting methods that elicit participants' self-consciousness about their daily behaviors at the minimum level.

Besides, several design opportunities visioned in this paper contain open questions for future work to explore. They call for coordinated efforts among CSCW researchers, technical experts, local governments, and migrant individuals and groups to work on the presented problem together.

## 8 CONCLUSION

HCI and CSCW research on crisis information seeking previously focused on domestic citizens of the affected regions. In this paper, we call attention to international migrants who need to survive a crisis in language and cultural environments that are different from their native ones. We presented findings from a qualitative study in the context of the COVID pandemic. Our participants consisted of two cohorts of Chinese migrants navigating COVID outbreak in Japan and the United States. In-depth interviews and self-tracking sessions were designed to understand participants' ongoing crisis information seeking behaviors, motivations, and concerns. We found that participants frequently visited Mandarin information resources to learn about COVID-related local information, such as breaking news, latest case number updates, and newly announced regulations and policies. This practice of taking a language detour of native language resources was triggered by multiple characteristics of the international migrants' life under COVID. It enhanced the migrants' perceived effectiveness of information gathering and sensemaking across multiple languages but exposed them to potential risks under high-stakes situations. Based on these findings, we discussed how digital technologies as well as social protocols could be designed to enhance people's navigation of multilingual information, access to and knowledge about local resources, and production of migrant-oriented content in the context of the COVID pandemic and beyond. Our research contributes to the future design of inclusive crisis infrastructures.

## ACKNOWLEDGMENTS

This work is supported by National Science Foundation, under grant 2147292 and grant 1753452. We thank Chi-Lan Yang, Mengyun Liu, and Young-Ho Kim for their assistance. We also thank Yongle Zhang, Yimin Xiao, Victoria Chang, and the anonymous reviewers for their valuable comments on earlier versions of this paper.

[34] Barney G Glaser. 1978. *Theoretical Sensitivity*. Mill Valley.

[35] Barney G Glaser and Anselm L Strauss. 2017. *The Discovery of Grounded Theory: Strategies for Qualitative Research*. Routledge.

[36] Google Translated Search. Retrieved from http://googlesystem.blogspot.com/2009/12/google-translated-search-nowmore.html

[37] Robert J Griffin, Sharon Dunwoody, and Kurt Neuwirth. 1999. Proposed model of the relationship of risk information seeking and processing to the development of preventive behaviors. *Environmental Research* 80, 2, 230–245.

[38] Robert J Griffin, Kurt Neuwirth, Sharon Dunwoody, and James Giese. 2004. Information sufficiency and risk communication. *Media Psychology* 6, 1, 23–61.

[39] Robert J Griffin, Kurt Neuwirth, James Giese, and Sharon Dunwoody. 2002. Linking the heuristic-systematic model and depth of processing. *Communication Research* 29, 6, 705–732.

[40] Lorenzo Guadagno. 2020. Migrants and the COVID-19 pandemic: An initial analysis. *Migration Research Series N° 60. International Organization for Migration (IOM)*. Geneva

[41] Xinning Gui, Yubo Kou, Kathleen Pine, Elisa Ladaw, Harold Kim, Eli Suzuki-Gill, and Yunan Chen. 2018. Multidimensional risk communication: Public discourse on risks during an emerging epidemic. *Proceedings of the 2018 CHI Conference on Human Factors in Computing Systems*. ACM, New York, NY, USA, 1–14.

[42] Xinning Gui, Yubo Kou, Kathleen H Pine, and Yunan Chen. 2017. Managing uncertainty: Using social media for risk assessment during a public health crisis. *Proceedings of the 2017 CHI Conference on Human Factors in Computing Systems*. ACM, New York, NY, USA, 4520–4533.

[43] Michael Hameleers, Toni GLA van der Meer, and Anna Brosius. 2020. Feeling "disinformed" lowers compliance with COVID-19 guidelines: Evidence from the US, UK, Netherlands, and Germany. *Harvard Kennedy School Misinformation Review* 1, 3.

[44] Stephen C Hines. 2001. Coping with uncertainties in advance care planning. *Journal of Communication* 51, 3, 498–513.

[45] Y Linlin Huang, Kate Starbird, Mania Orand, Stephanie A Stanek, and Heather T Pedersen. 2015. Connected through crisis: Emotional proximity and the spread of misinformation online. *Proceedings of the 18th ACM Conference on Computer Supported Cooperative Work and Social Computing*. ACM, New York, NY, USA, 969–980.

[46] Amanda L Hughes, Lise AA St. Denis, Leysia Palen, and Kenneth M Anderson. 2014. Online public communications by police & fire services during the 2012 Hurricane Sandy. *Proceedings of the SIGCHI Conference on Human Factors in Computing Systems*. ACM, New York, NY, USA, 1505–1514.

[47] Eslam Hussein, Prerna Juneja, and Tanushree Mitra. 2020. Measuring misinformation in video search platforms: An audit study on YouTube. *Proceedings of the ACM on Human-Computer Interaction* 4, CSCW 1, 1–27.

[48] Immuni. Retrieved from https://en.wikipedia.org/wiki/Immuni

[49] Interagency Language Roundtable. Retrieved from https://www.govtilr.org/Skills/ILRscale2.htm

[50] Tomoo Ito. 2020. A survey on the health of Vietnamese individuals living in Japan under a declared state of emergency due to COVID-19: A cross-sectional survey. *Research Square*.

[51] Shan Jiang and Christo Wilson. 2018. Linguistic signals under misinformation and fact-checking: Evidence from user comments on social media. *Proceedings of the ACM on Human-Computer Interaction* 2, CSCW, 1–23.

[52] Branden B Johnson. 2005. Testing and expanding a model of cognitive processing of risk information. *Risk Analysis: An International Journal* 25, 3, 631–650.

[53] Lee Ann Kahlor. 2007. An augmented risk information seeking model: The case of global warming. *Media Psychology* 10, 3, 414–435.

[54] Lee Ann Kahlor. 2010. PRISM: A planned risk information seeking model. *Health Communication* 25, 4, 345–356.

[55] Hye Kyung Kim, Jisoo Ahn, Lucy Atkinson, and Lee Ann Kahlor. 2020. Effects of COVID-19 misinformation on information seeking, avoidance, and processing: A multicountry comparative study. *Science Communication* 42, 5, 586–615.

[56] Young-Ho Kim, Jae Ho Jeon, Bongshin Lee, Eun Kyoung Choe, and Jinwook Seo. 2017. OmniTrack: A flexible self-tracking approach leveraging semi-automated tracking. *Proceedings of the ACM on Interactive, Mobile, Wearable and Ubiquitous Technologies* 1, 3 (2017), 1–28.

[57] Erica Kleinman, Sara Chojnacki, and Magy Seif El-Nasr. 2021. The gang's all here: How people used games to cope with COVID19 quarantine. *Proceedings of the 2021 CHI Conference on Human Factors in Computing Systems*. ACM, New York, NY, USA, 1–12.

[58] Marina Kogan and Leysia Palen. 2018. Conversations in the eye of the storm: At-scale features of conversational structure in a high-tempo, high-stakes microblogging environment. *Proceedings of the 2018 CHI Conference on Human Factors in Computing Systems*. ACM, New York, NY, USA, 1–13.

[59] Yubo Kou, Xinning Gui, Yunan Chen, and Kathleen Pine. 2017. Conspiracy talk on social media: Collective sensemaking during a public health crisis. *Proceedings of the ACM on Human-Computer Interaction* 1, CSCW, 1–21.